\documentclass[superscriptaddress,preprintnumbers,10pt,amsmath,amssymb,twocolumn,floats]{revtex4-1}
\usepackage{txfonts}
\usepackage{amssymb}
\usepackage{amsfonts}
\usepackage{amsmath}
\usepackage{graphicx}
\usepackage{epstopdf}
\usepackage{color}
\begin{document}
\title{Charge-density-wave-induced bands renormalization and energy gaps\\ in a kagome superconductor RbV$_3$Sb$_5$}

\author{Zhonghao Liu}
\email{lzh17@mail.sim.ac.cn}
\affiliation{State Key Laboratory of Functional Materials for Informatics, Shanghai Institute of Microsystem and Information Technology, Chinese Academy of Sciences, Shanghai 200050, China}
\affiliation{College of Materials Science and Opto-Electronic Technology, University of Chinese Academy of Sciences, Beijing 100049, China}

\author{Ningning Zhao}
\affiliation{Department of Physics and Beijing Key Laboratory of Opto-Electronic Functional Materials$\&$Micro-Nano Devices, Renmin University of China, Beijing 100872, China}

\author{Qiangwei Yin}
\affiliation{Department of Physics and Beijing Key Laboratory of Opto-Electronic Functional Materials$\&$Micro-Nano Devices, Renmin University of China, Beijing 100872, China}

\author{Chunsheng Gong}
\affiliation{Department of Physics and Beijing Key Laboratory of Opto-Electronic Functional Materials$\&$Micro-Nano Devices, Renmin University of China, Beijing 100872, China}

\author{Zhijun Tu}
\affiliation{Department of Physics and Beijing Key Laboratory of Opto-Electronic Functional Materials$\&$Micro-Nano Devices, Renmin University of China, Beijing 100872, China}

\author{Man Li}
\affiliation{Department of Physics and Beijing Key Laboratory of Opto-Electronic Functional Materials$\&$Micro-Nano Devices, Renmin University of China, Beijing 100872, China}

\author{Wenhua Song}
\affiliation{Department of Physics and Beijing Key Laboratory of Opto-Electronic Functional Materials$\&$Micro-Nano Devices, Renmin University of China, Beijing 100872, China}

\author{Zhengtai Liu}
\affiliation{State Key Laboratory of Functional Materials for Informatics, Shanghai Institute of Microsystem and Information Technology, Chinese Academy of Sciences, Shanghai 200050, China}

\author{Dawei Shen}
\affiliation{State Key Laboratory of Functional Materials for Informatics, Shanghai Institute of Microsystem and Information Technology, Chinese Academy of Sciences, Shanghai 200050, China}
\affiliation{College of Materials Science and Opto-Electronic Technology, University of Chinese Academy of Sciences, Beijing 100049, China}

\author{Yaobo Huang}
\affiliation{Shanghai Advanced Research Institute, Chinese Academy of Sciences, Shanghai 201204, China}

\author{Kai Liu}
\email{kliu@ruc.edu.cn}
\affiliation{Department of Physics and Beijing Key Laboratory of Opto-Electronic Functional Materials$\&$Micro-Nano Devices, Renmin University of China, Beijing 100872, China}

\author{Hechang Lei}
\email{hlei@ruc.edu.cn}
\affiliation{Department of Physics and Beijing Key Laboratory of Opto-Electronic Functional Materials$\&$Micro-Nano Devices, Renmin University of China, Beijing 100872, China}

\author{Shancai Wang}
\email{scw@ruc.edu.cn}
\affiliation{Department of Physics and Beijing Key Laboratory of Opto-Electronic Functional Materials$\&$Micro-Nano Devices, Renmin University of China, Beijing 100872, China}

\begin{abstract}
Recently discovered Z$_2$ topological kagome metals $A$V$_3$Sb$_5$ ($A$ = K, Rb, and Cs) exhibit charge density wave (CDW) phases and novel superconducting paring states, providing a versatile platform for studying the interplay between electron correlation and quantum orders. Here we directly visualize CDW-induced bands renormalization and energy gaps in RbV$_3$Sb$_5$ using angle-resolved photoemission spectroscopy, pointing to the key role of tuning van Hove singularities to the Fermi energy in mechanisms of ordering phases. Near the CDW transition temperature, the bands around the Brillouin zone (BZ) boundary are shifted to high-binding energy, forming an ``$M$"-shape band with singularities near the Fermi energy.  The Fermi surfaces are partially gapped and the electronic states on the residual ones should be possibly dedicated to the superconductivity. Our findings are significant in understanding CDW formation and its associated superconductivity.
\end{abstract}
\maketitle

\section{Introduction}
Layered kagome-lattice transition metals are emerging as an exciting platform to explore frustrated lattice geometry and quantum topology. A set of typical kagome-lattice electronic bands is produced by the tight-binding calculation featuring a Dirac dispersion at the Brillouin zone (BZ) corner, a saddle point at the zone boundary, and a flat band through the BZ \cite{CoSn_Liu}. Close-to-textbook kagome electronic bands with orbital differentiation physics have been experimentally observed in paramagnet CoSn \cite{CoSn_Liu}. In some kagome-lattice materials, the versatile quantum phenomena associating with the features near the Fermi energy ($E\rm_F$) have been found, such as Dirac/Weyl fermions \cite{Mn3Sn_Kuroda,Fe3Sn2_Linda,FeGeTe_Kim,CoSnS_LIU,CoSnS_Wang,CoMnGa_Hasan,CoSnS_YLChen,CoSnS_Beidenkopf,FeSn_Kang,FeSn_Lin,YMnSn_Li}, ferromagnetism \cite{Fe3Sn2_Lin,Fe3Sn2_Yin,FeGeTe_Zhang}, negative flat band magnetism \cite{CoSnS_Yin}, and topological Chern magnet \cite{TbMnSn_Yin}.

The theory was put forward early that a two-dimensional (2D) energy band with saddle points in the vicinity of $E\rm_F$ is unstable against charge density wave (CDW) formation \cite{Instability_Rice}. The CDW, superconducting and topological phases have been extensively investigated in 2D transition metal dichalcogenides \cite{TMD_Review} and the underlying microscopic mechanism of the CDW formation is still in controversy. Recently, the CDW state and superconductivity are discovered in a family of layered kagome metals $A$V$_3$Sb$_5$ ($A$ = K, Rb, and Cs) \cite{AV3Sb5_PRM_Toberer,AV3Sb5_PRL_Wilson,AV3Sb5_Sciadv_Mazhar,RbV3Sb5_HCLei,AV3Sb5_PRM_Wilson}, which hosts a Z$_2$ topological invariant and non-trivial topological Dirac surface states near $E\rm_F$  \cite{AV3Sb5_PRL_Wilson}.
The CDW state is probably driven by the competing electronic orders at the saddle-point singularity with a high density of states \cite{STM_Hasan,STM_Zeljkovic,STM_XHChen,STM_HJGao,Pressure_SYLi,Pressure_JGCheng,AV3Sb5_Optical,AV3Sb5_Josephson,AV3Sb5_HMiao,AV3Sb5_Cal_Yan,AV3Sb5_theory_Hu,kagome_theory_Li,kagome_theory_Wang,kagome_PRB_Thomale,kagome_PRL_Thomale}.  X-ray diffraction and scanning tunneling microscopy (STM) reveal the formation of a three-dimensional (3D) $2\times2\times2$ superlattice at both CDW and superconducting states \cite{AV3Sb5_PRM_Wilson,STM_Hasan,STM_Zeljkovic,STM_XHChen,STM_HJGao}, which energetically favors a chiral charge order and an inverse Star of David distortion in kagome lattice with the shift of van Hove singularity to $E\rm_F$ \cite{AV3Sb5_theory_Hu,AV3Sb5_theory_TNeupert,AV3Sb5_Cal_Yan}.  The CDW states and double superconducting domes are associated with multiple singularities with different energies and orbital characters near $E\rm_F$ \cite{STM_XHChen,STM_HJGao,Pressure_SYLi,Pressure_JGCheng,AV3Sb5_Optical,AV3Sb5_Josephson,CsV3Sb5_XLDong,AV3Sb5_HMiao,CsV3Sb5_HQYuan,CsV3Sb5_ZRYang,CsV3Sb5_XLChen,AV3Sb5_Cal_Yan,AV3Sb5_theory_Hu,kagome_theory_Li,kagome_theory_Wang}, which at present need to be further studied in detail.  In addition, a giant anomalous Hall effect with the reversal of Hall sign is observed \cite{AV3Sb5_Sciadv_Mazhar}, and magnetic order and local moment are not found by magnetic susceptibility and muon spin spectroscopy \cite{AV3Sb5_Sciadv_Mazhar,RbV3Sb5_HCLei,AV3Sb5_muon}.
To fundamentally understand these anomalous behaviors and quantum orders, the investigation of the temperature evolution of the low-energy electronic structure is highly desired.

\begin{figure*}[t!]
	\centering
	\includegraphics[width=0.84\textwidth]{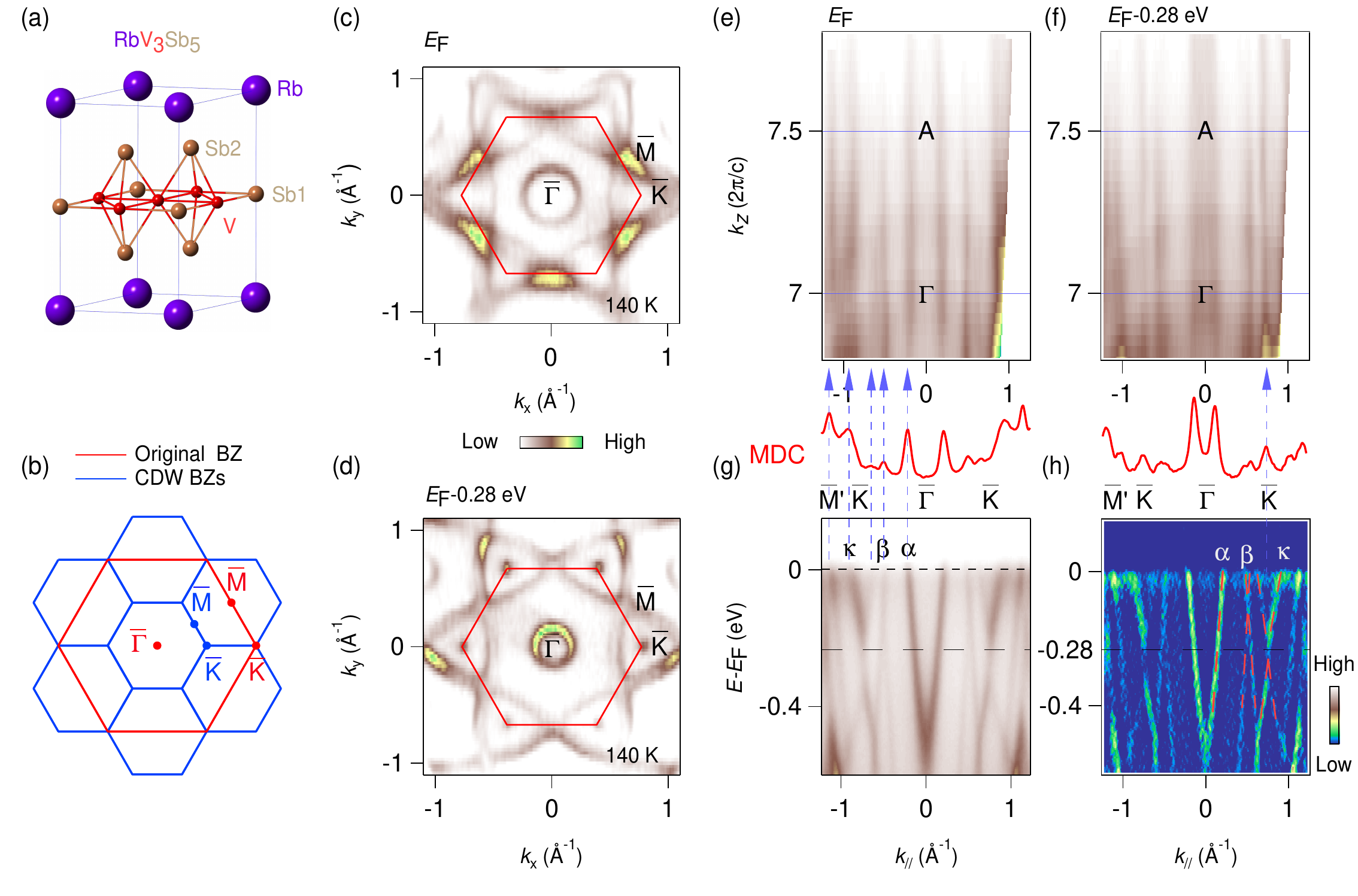}
	\caption
	{(a) Crystal structure of RbV$_3$Sb$_5$ with space group $P6/mmm$ (No. 191).
     (b) The original (red lines) and $2\times2$ reconstructed (blue lines) BZs projected on the (001) surface with the high-symmetry points.
     (c) and (d) Integrated intensity plot ($\pm$10 meV) at $E\rm_F$ and $E\rm_F$ - 0.28 eV taken at 140 K. The red lines indicate the high-symmetry directions and the original BZs.
     (e) and (f) Integrated intensity plots ($\pm$10 meV) on the $k_z$-$k_\parallel$ plane at $E\rm_F$ and $E\rm_F$ - 0.28 eV with $k_\parallel$ oriented along the $\bar{\Gamma}-\bar{K}$ direction. The high-symmetry points are plotted.
     (g) and (h) Intensity plot and corresponding second derivative plot along the $\bar{\Gamma}-\bar{K}$ direction. The MDCs taken at $E\rm_F$ are shown by the red curves. The bands are indicated by the Greek letters and the red dashed lines.
    }
	\label{1}
\end{figure*}

In this paper, we report on a combined angle-resolved photoemission spectroscopy (ARPES) and first-principles calculations study of the temperature evolution of the low-energy electronic structure in RbV$_3$Sb$_5$, which has a CDW transition temperature (T$\rm_{CDW}$) of about 102 K, a sign change of the Hall coefficient at about 40 K, and a superconducting transition temperature (T$\rm_c$) of about 0.92 K \cite{RbV3Sb5_HCLei}.  As a result of the CDW transition, we found that the bands at the zone boundary ($\bar{M}$) are shifted down about 40 meV forming an ``$M$"-shape band with its singularity at about 60 meV below $E\rm_F$. Below T$\rm_{CDW}$, the energy gap of about 20 meV is opened at the Fermi momentum ($k\rm_F$) of the band centered at $\bar{M}$, and no gap is observed at the band centered at $\bar{\Gamma}$ at 10 K within experimental energy resolution. The electronic states on the residual Fermi surfaces should be dedicated to the superconducting pairing. Our findings reveal CDW-induced strong bands renormalization and energy gaps at the zone boundary, implying that they are the multiple singularities at $M$ which play ultimate roles in the formation of both CDW and its related superconducting phases.

\section{Results and Discussion}
The crystal structure of RbV$_3$Sb$_5$ crystallizes in a hexagonal structure with $P6/mmm$ (No. 191) space group \cite{AV3Sb5_PRM_Toberer,AV3Sb5_PRL_Wilson,AV3Sb5_PRM_Wilson,RbV3Sb5_HCLei,AV3Sb5_Sciadv_Mazhar}, in which V-Sb slabs consisting of V kagome nets and interspersing Sb atoms are separated by alkali-metal ions along the $c$ axis, as shown in Fig. \ref{1}(a). There are two kinds of Sb sites: the Sb1 site at the centers of V hexagons, and the Sb2 site below and above the centers of V triangles forming hexagon layers. The corresponding original (red lines) and $2\times2$ reconstructed (blue lines) BZs projected on the (001) surface with the high-symmetry points are shown in Fig. \ref{1}(b). Figures \ref{1}(c) and \ref{1}(d) show constant-energy surfaces taken at 140 K at $E\rm_F$ and $E\rm_F$ - 0.28 eV, respectively.  The high intensity around the $\bar{M}$ points at the Fermi surfaces suggests the singularities or the surface states at the proximity of $E\rm_F$. To investigate the 3D character of the Fermi surfaces, we carried out the photon-energy-dependent ARPES measurement. With an empirical value of the inner potential of $\sim$ 8.2 eV and $c$ = 9.07 {\AA} \cite{AV3Sb5_PRM_Wilson}, we found that $hv$ = 86 eV is close to the $\Gamma$ point and 100 eV close to the $A$ point, according to the free-electron final-state model \cite{TlRbFeSe_Liu}. Three electronlike pockets ($\alpha$, $\beta$, and $\kappa$) along the  $\bar{\Gamma}-\bar{K}$ direction are indicated in Figs. \ref{1}(g) and \ref{1}(h). All of the three bands show weak $k_z$ dispersions both at $E\rm_F$ and $E\rm_F$ - 0.28 eV, as shown in Figs. \ref{1}(e) and \ref{1}(f), which reveal the 2D electronic dispersions along  $\bar{\Gamma}-\bar{K}$ and 2D Dirac cones at the $\bar{K}$ points.  Based on the ARPES data, we estimate that the widths of the $\alpha$, $\beta$, and $\kappa$ Fermi pockets along $\bar{\Gamma}-\bar{K}$ are about 0.42, 0.10, and 0.31  {\AA}$^{-1}$ and the Fermi velocities of them are about 3.25, 3.60, and 1.70 eV {\AA} (1.28 eV {\AA} for another branch of the Dirac bands), respectively.

\begin{figure*}[t!]
	\centering
	\includegraphics[width=0.71\textwidth]{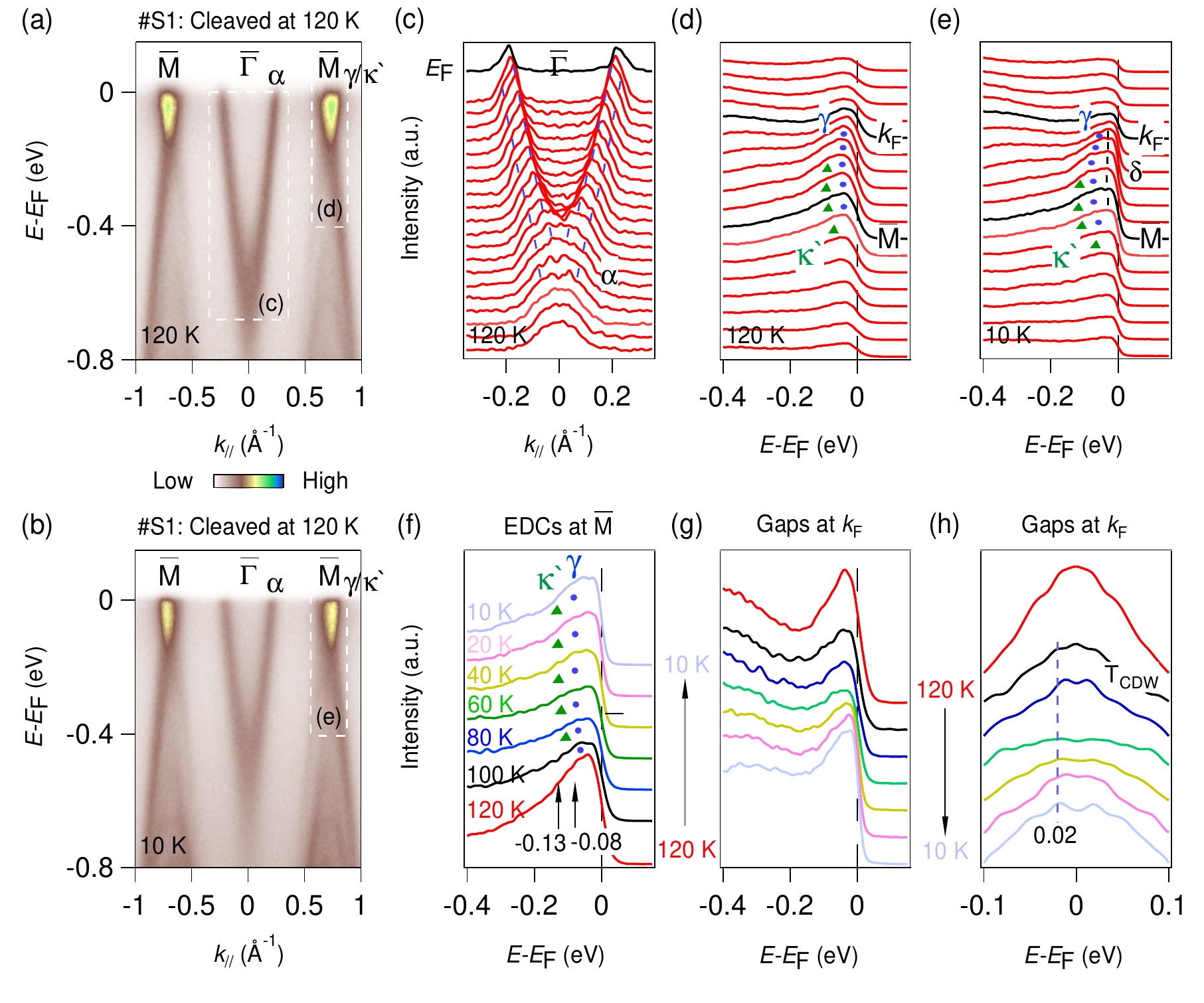}
	\caption
	{(a) and (b) Intensity plots along the $\bar{\Gamma}-\bar{M}$ direction taken at 120 and 10 K. The sample (\#S1) was cleaved at 120 K and measured along with the decreasing temperature. The bands are indicated by the Greek letters.
     (c) MDCs around the $\bar{\Gamma}$ point taken at 120 K, as indicated by the dashed rectangles in (a).
     (d) and (e) EDCs around the $\bar{M}$ point at 120 K and 10 K, as indicated by the dashed rectangles in (a) and (b). The $\kappa^\prime$ and $\gamma$ bands are indicated by different color makers.
     (f) EDCs at the $\bar{M}$ center taken at different temperatures.
     (g) and (h) EDCs and their symmetrizations at the $k\rm_F$ of the $\gamma$ band along $\bar{\Gamma}-\bar{M}$ taken at different temperatures. Different colors represent different temperatures.
     }
	\label{2}
\end{figure*}

Figure \ref{2} shows the temperature evolution of the bands along the $\bar{\Gamma}-\bar{M}$ direction on a sample cleaved at 120 K (\#S1). The intensity plots along the $\bar{\Gamma}-\bar{M}$ direction taken at 120 and 10 K are shown in Figs. \ref{2}(a) and \ref{2}(b), respectively. Comparing the data taken at the two temperatures, one can see that the $\alpha$ band is shifted up, which is mainly attributed by surface reconstructions along with time \cite{CsV3Sb5_JFHe}. Figure \ref{2}(c) shows the momentum distribution curves (MDCs) of the $\alpha$ band taken at 120 K, revealing the two splitting sub-branches. The STM results suggest an isotropic scattering vector connecting different states of the $\alpha$ pocket \cite{STM_Zeljkovic,STM_XHChen,STM_HJGao}, while the two Sb sites or $k_z$ integration can also cause the bands splitting in the ARPES data. We have carried out substantial experiments on the samples with various conditions, \emph{e.g.} cleaved at both high and low temperatures then measured them at a few stabilized temperatures along with decreasing/increasing temperatures, as shown in the supplementary Fig. S(1) \cite{SM_here}. Our data reveal that the $\alpha$ band at $\Gamma$ is sensitive to the sample surface and the vacuum, while the other bands are not.

At the $\bar{M}$ point, the $\kappa^\prime$  and $\gamma$ bands can be observed at 120 K [Fig. \ref{2}(d)]. The two bands are shifted down at 10 K and the $\delta$ band is brought out [Fig. \ref{2}(e)]. The $\delta$ band could be an edge state cut by the Fermi distribution function.  We estimate that the widths of the $\alpha$, $\kappa^\prime$, and $\gamma$ Fermi pockets along $\bar{\Gamma}-\bar{M}$ at 10 K are about 0.42, 0.22, and 0.10 {\AA}$^{-1}$ and the Fermi velocities of them are about 3.32, 1.68, and 4.20 eV {\AA}, respectively. We display in detail the temperature-dependent data in Figs. \ref{2}(f)-\ref{2}(h). As shown in Fig. \ref{2}(f), one can find that the $\kappa^\prime$ and $\gamma$ bands begin to be shifted down around T$\rm_{CDW}$ (100 K) and stand steadily around 60-80 K with their band bottoms at 0.08 and 0.13 eV below $E\rm_F$, respectively. Below $\sim$ 80 K, the $\gamma$ band further opens the energy gap of about 20 meV, as shown in the energy distribution curves (EDCs) and their symmetrizations at $k\rm_F$ [Figs. \ref{2}(g) and \ref{2}(h)]. While we do not observe the CDW gap at the $\alpha$ band centered at $\bar{\Gamma}$ at low temperatures along both $\bar{\Gamma}-\bar{M}$ and $\bar{\Gamma}-\bar{K}$, as shown in the supplementary Fig. S(2) \cite{SM_here}.

\begin{figure*}[t!]
	\centering
	\includegraphics[width=0.95\textwidth]{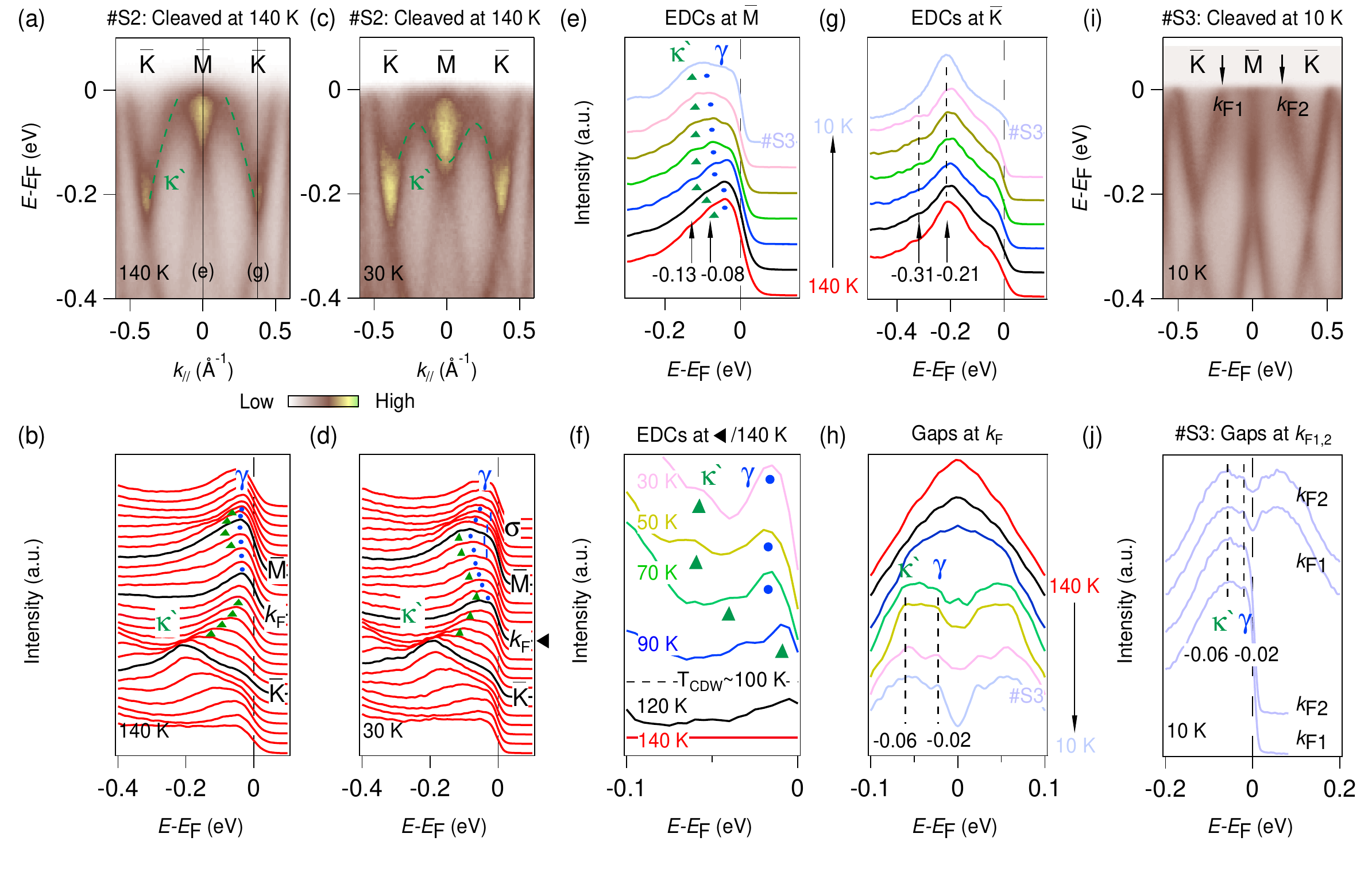}
	\caption
	{(a)-(d) Intensity plots and corresponding EDCs along the $\bar{K}-\bar{M}$ direction taken at 140 and 30 K, respectively. The sample (\#S2) was cleaved at 140 K and
     measured along with the decreasing temperature. The $\kappa^\prime$ and $\gamma$ bands are indicated by different color makers.
     (e) EDCs at the $\bar{M}$ center as indicated by the line in (a), taken at different temperatures. The energy positions are indicated by the black arrows.
     (f) EDCs at the fixed momentum [indicated by $\blacktriangleleft$ in (d)] taken at different temperatures and divided by the EDC taken at 140 K.
     (g) EDCs at the $\bar{K}$ center as indicated by the line in (a), taken at different temperatures.
     The energy positions are indicated by the black arrows.
     (h) The symmetrized EDCs at the $k\rm_F$ of the $\gamma$ band along $\bar{K}-\bar{M}$ taken at different temperatures. The values of gaps are marked.
     \#S3 represents the EDCs taken on the third sample.
     (i) Intensity plot along the $\bar{K}-\bar{M}$ direction taken on the freshly cleaved sample at 10 K (\#S3). $k\rm_F$ is indicated by the black arrows.
     (j) EDCs and their symmetrizations at the $k\rm_F$ of the $\gamma$ band along $\bar{K}-\bar{M}$, as indicated in (i).
    }
	\label{3}
\end{figure*}

\begin{figure*}[t!]
	\centering
	\includegraphics[width=0.95\textwidth]{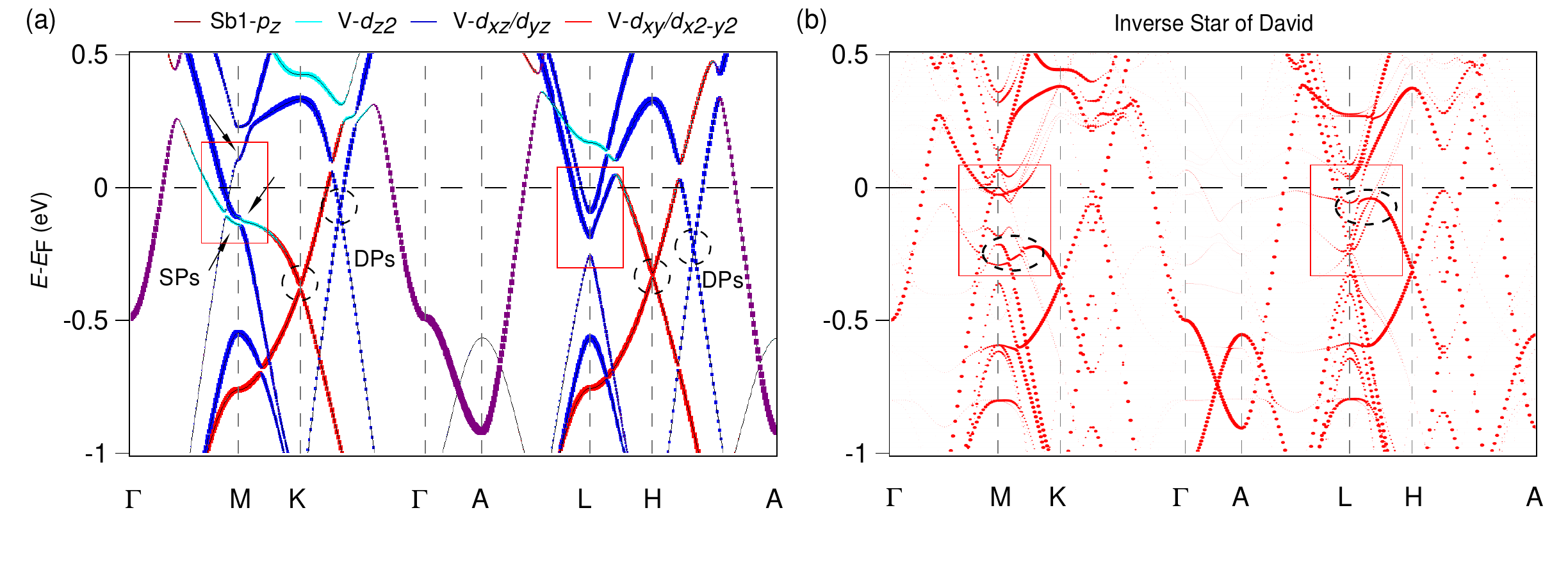}
	\caption
	{(a) Orbital-projection band structure calculation of RbV$_3$Sb$_5$ with SOC for the normal state. Here the main contribution orbitals near $E\rm_F$ are shown, and the orbitals weights are represented by both the colors and the size of the bands. The orbitals weights of V atoms are the average values of the adjacent three V atoms. The saddle points (SPs) are indicated by the arrows, and the Dirac points (DPs) are marked with the dashed circles. The features at $\bar{M}$ ($M$ and $L$) are indicated by the red squares.
  (b) The unfolded band structure of the inverse Star of David phase. The bands with a strong renormalization at the $M$ and $L$ points are marked with the dashed ellipses. The marked band at $L$ sinks below $E\rm_F$ in the CDW state.
    }
	\label{4}
\end{figure*}

Figures \ref{3}(a)-\ref{3}(h) show the temperature evolution of the bands along the $\bar{K}-\bar{M}$ direction on a sample cleaved at 140 K (\#S2). From the intensity plots and the corresponding EDCs, one can clearly see that the bands near the $\bar{M}$ point are remarkably renormalized by CDW. The $\kappa^\prime$ band crosses $E\rm_F$ at 140 K as shown in Figs. \ref{3}(a) and \ref{3}(b). The $\kappa^\prime$ and $\gamma$ bands shifting down along with decreasing temperature mentioned above can be more clearly identified along the $\bar{K}-\bar{M}$ direction, as shown in Figs. \ref{3}(a)-\ref{3}(e). More strikingly, the $\kappa^\prime$ band centered at $\bar{M}$ is flatten and sunk below $E\rm_F$ at $\sim$ 50-70 K, as shown in Figs. \ref{3}(d) and \ref{3}(f), forming an ``$M$"-shape band with the tips of the $\kappa^\prime$ band (singularities) at about 60 meV below $E\rm_F$. The $\gamma$ band along $\bar{K}-\bar{M}$ is also further opened the energy gap of about 20 meV, as shown in the symmetrized EDCs at the $k\rm_F$ of Figs. \ref{3}(h). The temperature evolution of the $\kappa^\prime$ band seems like the calculation with an inverse Star of David pattern in Fig. \ref{4}(b).

To check the CDW-induced bands renormalization, we directly compare the data taken on the freshly cleaved samples at low temperature (\#S6: Cleaved at 10 K) and high temperature (\#S7: Cleaved at 140 K), as shown in the supplementary Fig. S(3) \cite{SM_here}. The sharp contrast between Figs. S(3)(a) and S(3)(b) reveals that the band renormalization is indeed induced by temperature rather than a trivial surface reconstruction. We also measure the energy gaps on a freshly cleaved sample at 10 K (\#S3), as shown in Figs. \ref{3}(i) and \ref{3}(j). The energy positions in Fig. \ref{3}(j) show CDW gaps at -20 meV and the tip of $\kappa^\prime$ band at about -60 meV, respectively. In addition, the temperature evolution of the Dirac cone at the $\bar{K}$ point is shown in Fig. \ref{3}(g). The Dirac point is almost not moved along with decreasing temperature. The energy gap of the Dirac cone is about 100 meV, which is much larger than the calculated value of $\sim$ 15-25 meV induced by spin-orbit interaction in the normal state [Fig. \ref{4}(a)] and is close to the calculated value of $\sim$ 120 meV in the CDW state [Fig. \ref{4}(b)]. The CDW-gaps induced by the band folding at the Dirac cones are needed to be considered.

With the help of the orbital-projection band calculation [Fig. \ref{4}(a) and the supplementary Figs. S(4) \cite{SM_here}], one can find the $\alpha$ band at $\bar{\Gamma}$ is mainly contributed by the out-of-plane Sb1-$p_z$ (dark), and the $\kappa^\prime$ and $\gamma$  bands at $\bar{M}$ are mainly derived from the out-of-plane V-$d_{z2}$ (green) and V-$d_{xz}$/$d_{yz}$ (blue) orbitals. The bonding of the out-of-plane orbitals and the interlayer coupling strength are enhanced along with the decreasing of the temperature, which is also revealed by the reduction of the $c$-axis lattice constant \cite{AV3Sb5_PRL_Wilson}. The renormalization of the bands with the out-of-plane character should be more appreciated with the CDW instability. Coulomb scattering of electrons between the orbital-selective saddle-point singularities at $M$ can give rise to instabilities of the Fermi surfaces and lead to CDW states \cite{STM_Hasan,STM_Zeljkovic,STM_XHChen,STM_HJGao,Pressure_SYLi,Pressure_JGCheng,AV3Sb5_Optical,AV3Sb5_Josephson,AV3Sb5_HMiao,AV3Sb5_Cal_Yan,AV3Sb5_theory_Hu,kagome_theory_Li,kagome_theory_Wang,kagome_PRB_Thomale,kagome_PRL_Thomale}. The 2D Dirac bands at $\bar{K}$ ($\kappa$ and $\kappa^\prime$) originate from in-plane V-$d_{xy}$/$d_{x2-y2}$ (red) orbitals, and hybridized with out-of-plane V-$d_{xz}$/$d_{yz}$ (blue) orbitals near the $\bar{M}$ point. The Dirac bands remain nearly motionless upon the CDW phase transition.

The calculation in CDW state with the inverse Star of David can well describe our experimental observations [Fig. \ref{4}(b) and the supplementary Fig. S(5) \cite{SM_here}]. As marked with the dashed ellipses in Fig. \ref{4}(b), the $\kappa^\prime$ band at the $L$ point sinks below $E\rm_F$, forming an ``$M$"-shape band with its tips at about 60 meV. The tips can be viewed as new singularities, which may be further associated with superconducting states. The CDW-induced band renormalization is endowed with an electronic correlation effect. Previous studies provide strong evidence that traversing the singularity to $E\rm_F$ is beneficial in the formation of ordering phenomena. For instance, in CDW-material TaSe$_2$ also with 3D $2\times2\times2$ superstructure, by tuning the energy position of the singularity, the T$\rm_{C}$ is enhanced by more than an order of magnitude \cite{TaSe2_SVB}. Recently, thermal conductivity and high-pressure resistance measurements reveal two superconducting domes and exotic pairing states \cite{Pressure_SYLi,Pressure_JGCheng,CsV3Sb5_ZRYang,CsV3Sb5_XLChen}, which may be associated with the optimal positions of the singularities concerning $E\rm_F$ match with corresponding bosons.

Besides the bands renormalization mentioned above, the CDW-induced energy gap $\Delta\sim$ 20 meV is opened at $k\rm_F$ of the band near $\bar{M}$ which is consistent with STM results \cite{STM_Hasan,STM_Zeljkovic,STM_XHChen,STM_HJGao}. The CDW-induced gap is not observed at the band near $\bar{\Gamma}$ at T = 10 K within the experimental energy resolution. Thus, momentum- and orbital-dependence of the electronic states are involved in the CDW formation in RbV$_3$Sb$_5$. The STM measurements further reveal that the CDW gap is particle-hole asymmetric \cite{STM_Hasan,STM_Zeljkovic,STM_XHChen,STM_HJGao}, which is previously found in CDW-material NbSe$_2$  \cite{NbSe2_Hoffman,NbSe2_SRosenkranz}. As a case of typical quasi-2D materials without a strongly nested Fermi surface, the presence of a particle-hole asymmetric gap in NbSe$_2$ could be an indication that electron correlation is important in driving the CDW  \cite{NbSe2_Hoffman,NbSe2_SRosenkranz}. Analogously, combined with the large ratio 2$\Delta$/$k\rm_{B}$T$\rm_{CDW}$ $\sim$ 4.55 in analogy to strong-coupling superconductors, the CDW formation in RbV$_3$Sb$_5$ is likely mediated by electronic interactions enhanced by low dimensionality. Recent inelastic x-ray scattering studies demonstrate an unconventional and electronic-driven mechanism that couples the CDW and the topological band structure in RbV$_3$Sb$_5$ \cite{AV3Sb5_HMiao}.

In addition, as a Z$_2$ topological kagome metal, RbV$_3$Sb$_5$ hosts non-trivial topological Dirac surface states at the time-reversal-invariant $\bar{M}$ points and remains the same after the CDW transition, as shown in the supplementary Fig. S(6) \cite{SM_here}. It is possible to realize the Majorana zero-energy modes and their related topological superconductivity in these materials. Because the electronlike bands near the $\bar{M}$ point show $k_z$ dispersions [the supplementary Fig. S(7) \cite{SM_here}], the surface states at $\bar{M}$ are possibly located above $E\rm_F$. The chemical potential need to be elevated for further studying the surface states in detail.

\section{Conclusion}
In summary, we have studied the electronic structures of a kagome superconductor RbV$_3$Sb$_5$ in both the normal phase and the CDW phase. We observed the CDW-induced bands renormalization and energy gaps on the bands at the zone boundary, where multiple orbital-selective singularities exist. Momentum- and orbital-dependence of the electronic states are involved in the CDW formation and the associated superconductivity. Our findings strongly imply that the singularities near $E\rm_F$ play important roles in the formation of ordering phases and the electronic states on the residual Fermi surfaces to the superconducting pairing.

\section{Acknowledgements}
This work was supported by the National Key R$\&$D Program of the Ministry of Science and Technology of China (MOST) (Grants No. 2018YFE0202600, 2017YFA0302903, and 2016YFA0300504), the National Natural Science Foundation of China (NSFC) (Grants No. 11822412, 11774421, 11774423, and 11774424), the Beijing Natural Science Foundation (Grant No. Z200005), the CAS Interdisciplinary Innovation Team, the Fundamental Research Funds for the Central Universities, the Research Funds of Renmin University of China (RUC) (Grants No. 18XNLG14, 19XNLG03, and 19XNLG17), and the Beijing National Laboratory for Condensed Matter Physics. N.Z. was supported by the Outstanding Innovative Talents Cultivation Funded Programs 2021 of RUC. The ARPES experiments were performed on the Dreamline beamline of SSRF and supported by MOST (Grant No. 2016YFA0401002), and the 03U Beamline of the SSRF is supported by the ME2 project (Grant No. 11227902) from NSFC. Computational resources were provided by the Physical Laboratory of High-Performance Computing at RUC and Shanghai Supercomputer Center.

Z.L., K.L., H.L., and S.W. provided strategy and advice for the research. Z.L., M.L., W.S., Z.L., D.S., Y.H., and S.W. performed ARPES measurements;  N.Z. and K.L. carried out the calculations; Q.Y., C.G., Z.T., and H.L. synthesized the single crystals. All authors contributed to writing the manuscript. Z.L., N.Z., and Q.Y. contributed equally to this work.

\section{APPENDIX: Methods}
Single crystals of RbV$_3$Sb$_5$ were synthesized by the self-flux method as described elsewhere \cite{RbV3Sb5_HCLei}. RbV$_3$Sb$_5$ single crystals are stable in the air. ARPES measurements were performed at the Dreamline and 03U beamlines of the Shanghai Synchrotron Radiation Facility (SSRF). The energy and angular resolutions were set to 10-24 meV and 0.02 {\AA}$^{-1}$, respectively.  The Fermi cut-off of the samples was referenced to an evaporated gold film on the sample holder.  Samples were cleaved $in$ $situ$, exposing flat mirrorlike (001) surfaces. The pressure was maintained at less than $2\times10^{-10}$ Torr during temperature-dependent measurements.

The first-principles electronic structure calculations on RbV$_3$Sb$_5$ were performed by using the projector augmented wave (PAW) method \cite{PAW_PRB_Blochl,PAW_PRB_Joubert} as implemented in the Vienna $ab$ $initio$ simulation package (VASP) \cite{VASP}. The generalized gradient approximation (GGA) of Perdew-Burke-Ernzerhof (PBE) type \cite{GGA} was used for the exchange-correlation functional. The kinetic energy cutoff of the plane-wave basis was set to be 350 eV. The BZ was sampled with a $10\times10\times6$ $k$-point mesh. For the Fermi surface broadening, the Gaussian smearing method with a width of 0.05 eV was adopted. The zero-damping DFT-D3 method was adopted to describe the interlayer van der Waals (vdW) interactions \cite{DFT_D}. The lattice constants and the atomic positions were fully relaxed until the forces on all atoms were smaller than 0.01 eV/{\AA}. The relaxed lattice constants $a$ = $b$ = 5.4333 {\AA} and $c$ = 8.9986 {\AA} are consistent with the experimental result \cite{AV3Sb5_PRM_Wilson}. The surface states in the projected 2D BZ were calculated with the surface Green's function method by using the WannierTools package \cite{Wannier_Soluyanov}. The tight-binding Hamiltonian of the semi-infinite system was constructed by the maximally localized Wannier functions \cite{Wannier_Marzari}. To study the CDW phase of RbV$_3$Sb$_5$, a $2\times2\times1$ supercell and a $5\times5\times5$  k-point mesh for the corresponding BZ sampling were used. The initial atomic distortions were firstly set according to the in-plane structures of the previously reported Star of David and inverse Star of David patterns \cite{AV3Sb5_Cal_Yan}, and then both the lattice parameters and the internal atomic positions were fully relaxed. The band structures of the CDW phases were unfolded in the BZ of the unit cell with the band unfolding method \cite{Supercell_AZunger} as in the PyVaspwfc package \cite{PyVaspwfc}.

\bibliography{Rb135}
\end{document}